\documentclass{sf2a-conf2012}
\usepackage{graphicx}
\usepackage{hyperref}
\usepackage[]{natbib}  
\usepackage{epstopdf}

\def\BibTeX{{\rm B\kern-.05em{\sc i\kern-.025em b}\kern-.08em
    T\kern-.1667em\lower.7ex\hbox{E}\kern-.125emX}}
\bibpunct{(}{)}{;}{a}{}{,}  
\newcommand{\teff}{$T_{\rm eff}$}
\newcommand{\logg}{$\log g$}
\newcommand{\ag}{$A_{\rm G}$}
\newcommand{\mlsep}{$\langle \Delta \nu \rangle$}

\newcommand{\numax}{$\nu_{\rm max}$}
\newcommand{\pmm}{$\pm$}
\newcommand{\msol}{M$_{\odot}$}
\newcommand{\rsol}{R$_{\odot}$}

\newcommand{\kms}{kms$^{-1}$}
\newcommand{\rad}{$R$}
\newcommand{\mass}{$M$}
\newcommand{\lum}{$L$}
\newcommand{\age}{$t$}

\newcommand{\mhz}{$\mu$Hz}
\newcommand{\feh}{[Fe/H]}
\newcommand{\kep}{{\it Kepler}}
\newcommand{\corot}{CoRoT}
\newcommand{\gspp}{{\tt GSP\_Phot}}

\begin{document}

\TitreGlobal{SF2A 2012}


\title{Asteroseismic constraints for Gaia}

\runningtitle{Asteroseismic constraints for Gaia}

\author{O. L. Creevey}\address{Laboratoire Lagrange, CNRS, Universit\'e de 
Nice Sophia-Antipolis, Nice, 06300, France}

\author{F. Th\'evenin$^1$}




\setcounter{page}{237}


\maketitle


\begin{abstract}
Distances from the Gaia mission will no doubt improve our understanding 
of stellar physics by providing an excellent constraint on the luminosity 
of the star.
However, it is also clear that high precision stellar properties from, 
for example, asteroseismology, will also provide a needed input constraint 
in order to calibrate the methods that Gaia will use, e.g. stellar models or 
\gspp.
For solar-like stars (F, G, K IV/V), asteroseismic data delivers 
at the least two very important quantities:
(1) the average large frequency separation $\langle \Delta \nu \rangle$ and 
(2) the frequency 
corresponding to the maximum of the modulated-amplitude spectrum 
$\nu_{\rm max}$.
Both of these quantities are related directly to stellar parameters 
(radius and mass) and in particular their combination (gravity and density).
We show how the precision in $\langle \Delta \nu \rangle$, $\nu_{\rm max}$, and 
atmospheric 
parameters $T_{\rm eff}$ and [Fe/H] affect the determination of gravity 
($\log g$) for a sample of well-known stars.
We find that $\log g$ can be determined within less than 0.02 dex accuracy
for our sample while considering precisions in the data expected for 
$V\sim12$ stars from {\it Kepler} data.
We also derive masses and radii which are accurate to within 1$\sigma$ of 
the accepted values.
This study validates the subsequent use of all of the available 
asteroseismic data on solar-like stars from the {\it Kepler} field 
($>500$ IV/V stars) in order to provide a very important 
constraint for Gaia calibration of \gspp\ through the use of $\log g$.
We note that while we concentrate on IV/V stars, 
both the CoRoT and {\it Kepler} fields contain asteroseismic 
data on thousands of giant stars which will also provide 
useful calibration measures.

\end{abstract}



\section{Introduction}

The ESA Gaia\footnote{http://sci.esa.int/science-e/www/area/index.cfm?fareaid=26} 
mission is due to launch in Autumn 2013.
Its primary objective is to perform a 6-D mapping of the Galaxy
by observing over 1 billion stars down to a magnitude of $V = 20$.
The mission will yield distances to these stars, and 
for about 20/100 million stars, distances with precisions of less than
1\%/10\% will be obtained.

Gaia will obtain its astrometry by using broad band ``G'' photometry 
(similar to a $V$ magnitude).
The spacecraft is also equipped with a spectrophotometer comprising both
a blue and a red prism BP/RP, delivering {\it colour} information.
A spectrometer will be used to determine the 
radial velocities of objects as far as $G=17$ (precisions from
1--20 \kms), and for stars with $G<11$ high resolution spectra 
(R$\sim$11,500) will be available.

One of the main workpackages devoted to source characterisation is 
{\gspp} whose
objectives are to obtain stellar properties for 1 billion single stars
by using 
the $G$ band photometry, the parallax $\pi$, and the spectrophotometric
information BP/RP \citep{bj10ilium}.
The stellar properties that will be derived are 
effective temperature \teff, extinction \ag\ in
the G band, surface gravity \logg, and metallicity \feh.
\citet{liu12} compare different methods to
determine these parameters and they
estimate typical precisions in \logg\ on the order
of 0.1 - 0.2 dex for main sequence late-type stars, and 
mean absolute residuals (true value minus inferred value from simulations) 
no less than 0.1 dex for stars of all magnitudes.

A calibration plan using forty bright benchmark stars has been put in
place to deliver the {\it best} stellar models.
These will be used on $\sim$5000 calibrations stars which will be observed
by Gaia.  However, for most of these fainter stars \logg\ remains quite
unconstrained, and this will inherently reduce the full capacity of 
source characterisation with Gaia data.

In the last decade or so, much progress in the field of 
observational asteroseismology
has been made, especially for stars exhibiting Sun-like 
oscillations.
These stars have deep outer convective envelopes where stochastic turbulence
gives rise to a broad spectrum of excited resonant oscillation modes 
e.g. \citealt{bg94}.  
The power spectra of such stars can be characterised by two mean seismic
quantities: \mlsep\ and \numax.
The quantity \mlsep\ is the mean value
of the {\it large frequency separations} 
$\Delta\nu_{l,n} = \nu_{l,n} - \nu_{l,n-1}$ where
$\nu_{l,n}$ is a frequency with degree $l$ and radial
order $n$, and \numax\ is the frequency corresponding to the maximum amplitude
of the bell-shaped frequency spectrum.
The following scaling relations have also been shown to hold:
Eq.~1: $\langle \Delta \nu \rangle
\approx M^{0.5}R^{-1.5}\langle \Delta \nu \rangle_{\odot}$
and 
Eq.~2: $\nu_{\rm max} \approx M R^{-2} (T_{\rm eff}/5777)^{-0.5}\nu_{\rm max, \odot}$ (Eq. 1)
where $\langle \Delta \nu \rangle_{\odot} = 134.9$ \mhz\ and 
$\nu_{\rm max, \odot} = 3,050$ \mhz\ \citep{kb95}.

Of particular interest for Gaia is the {\it Kepler} ({\url{http://kepler.nasa.gov}) field of view --- $\sim$100
square-degrees, centered on galactic coordinates 
76.32$^{\circ}$, +13.$5^{\circ}$.
{\it Kepler} is a NASA mission dedicated to characterising planet-habitability 
\citep{bor10science}.
It obtains photometric data of $\sim$150,000 stars with a typical 
cadence of 30 minutes.  A subset of stars ($<$ 1000 every 
month) acquire data with a point every 1 minute.  
This is sufficient to detect and characterise Sun-like oscillations in many stars.
\citet{ver11} and \citet{cha11science} recently showed the detections
of these mean seismic quantities for a sample of $>$500 F, G, K 
IV/V stars with typical magnitudes $7<V<12$, 
while both \corot\ and \kep\ have both shown their capabilities of 
detecting these
same seismic quantities in 1000s of red giants \citep{hek09, bau11, mos12}.

With the detection of mean seismic quantities in hundreds of stars,
the {\it Kepler} field is very promising for helping to calibrate the 
{\tt \gspp} methods.  In particular, they deliver
one of the four stellar properties to be extracted by automatic analyses
of Gaia data, namely
\logg. 
\citet{gai11} studied the distribution of errors for a 
sample of simulated stars using seismic data and a grid-based method 
based on stellar evolution models.  They concluded that a 
seismic \logg\ is almost fully independent of the input physics in the 
stellar evolution models that are used.
More recently \citet{mm11} compared classical determinations of \logg\
to those derived alone from the scaling relation (Eq. [2]), 
and concluded 
that the mean differences between the various methods used 
is $\sim$0.05 dex, thus supporting the validity of a 
seismic determination of \logg. 
However, to date, no study has been done to validate the {\it accuracy} of 
a seismic \logg\ (how closely it resembles the true value)
by using stars with measured radii and masses.
This is the objective of this work.



\section{A comparison of the direct and seismic methods for determining \logg.
\label{sec:section3}} 

\subsection{Observations and direct determination of \logg}

We aim to compare an {\it asteroseismically} derived \logg\ with the 
true known value for a sample of stars.
We chose a sample of seven bright well-characterised stars for which the 
radius is known via interferometry or a binary solution and the 
mass is known from either the binary solution or a detailed seismic analysis.
Table~\ref{tab:refstars} lists the sample of stars along with the observed
values of \mlsep, \numax, \teff, \feh, $M$, and $R$.  
The final column in the table gives the {\it true} value of \logg\ derived
from $M$ and $R$.   

\begin{table*}
\begin{center}
\caption{Observed properties of the reference stars\label{tab:refstars}}
\begin{tabular}{lccccccccccccccccccc}
\hline\hline
Star & \mlsep & \numax & \teff\ & \feh & \rad & \mass & \logg\\
&($\mu$Hz) & (mHz) & (K) & (dex) & (\rsol)&(\msol)&(dex)\\
\hline
$\alpha$CenB &161.5\pmm0.11$^{1a}$&4.0$^{1a}$&5316\pmm 28$^{1b}$&0.25\pmm 0.04$^{1b}$&0.863\pmm 0.005$^{1c}$ &0.934\pmm 0.0061$^{1d}$&4.538\pmm 0.008\\
18 Sco & 134.4\pmm 0.3$^{2a}$ & 3.1$^{2a}$ & 5813\pmm 21$^{2a}$ & 0.04\pmm 0.01$^{2a}$ & 1.010\pmm 0.009$^{2a}$ & 1.02\pmm 0.03$^{2a}$ & 4.438\pmm 0.005\\
Sun & 134.9\pmm 0.1$^{3a}$ & 3.05$^{3b}$ & 5778\pmm 20$^{3c}$ & 0.00\pmm 0.01$^{3d}$ & 1.000\pmm1.010$^{3d}$ & 1.000\pmm0.010$^{3d}$ & 4.438\pmm 0.002\\
$\alpha$CenA&105.6$^{4a}$&2.3$^{4a}$&5847\pmm 27$^{1b}$& 0.24\pmm 0.03$^{1b}$&1.224\pmm 0.003$^{1c}$&1.105\pmm 0.007$^{1d}$&4.307\pmm 0.005 \\
HD\,49933 &85.66\pmm 0.18$^{5a}$&1.8$^{5a}$&6500\pmm 75$^{5b}$ &-0.35\pmm 0.10$^{5b}$ 
&1.42\pmm 0.04$^{5c}$ & 1.20\pmm 0.08$^{5c}$ & 4.212\pmm 0.039\\
Procyon & 55.5\pmm 0.5$^{6a}$ & 1.0$^{6b}$ & 6530\pmm 90$^{6c}$ & -0.05\pmm 0.03$^{6d}$ & 2.067\pmm 0.028$^{6e}$ & 1.497\pmm 0.037$^{6f}$ & 3.982\pmm 0.016\\
$\beta$Hydri&57.24\pmm 0.16$^{7a}$ &1.0$^{7a}$&5872\pmm 44$^{7b}$ &-0.10\pmm 0.07$^{7c}$ & 1.814\pmm 0.017$^{7b}$ & 1.07\pmm 0.03$^{7b}$ & 3.950\pmm 0.015\\
\hline\hline
\end{tabular}
\end{center}
\begin{small}
References:
$^{1a}$\citet{kje05}, $^{1b}$\citet{pdm08}, $^{1c}$\citet{ker03}, $^{1d}$\citet{pou02},
$^{2a}$\citet{baz11},
$^{3a}$Taking the average of Table~3 from \citet{tf92}, $^{3b}$\citet{kb95}, $^{3c}$\citet{gs98}, $^{3d}$We adopt a typical error of 0.01 in \feh, $M$ and $R$, 
$^{4a}$\citet{bc02}, 
$^{5a}$Using the $l=0$ modes with Height/Noise$>$1 from Table 1 of \citet{ben09}, $^{5b}$\citet{kal10} $Z=0.008 \pm 0.002$ is referenced, 
$^{5c}$\citet{big11}, 
$^{6a}$\citet{egg04}, $^{6b}$\citet{mar04}, $^{6c}$\citet{fuh97}, $^{6d}$\citet{all02}, $^{6e}$\citet{ker04procyon}, $^{6f}$\citet{gir00},
$^{7a}$\citet{bed07}, $^{7b}$\citet{nor07}, $^{7c}$\citet{bru10}. 
\end{small}
\end{table*}


\subsection{Seismic method to determine \logg\label{sec:radex10}}
We use a grid-based method, RadEx10, to determine an asteroseismic value 
of \logg\ \citep{cre12kep}. 
The grid was constructed using the  ASTEC stellar evolution code 
\citep{jcd08astec} without diffusion effects and the same input physics as
described in \citet{cre12kep}.

The grid considers models with masses $M$ from 0.75 -- 2.0 \msol\ in steps of 
0.05 \msol, ages $t$ from ZAMS to subgiant, the initial chemical composition 
$Z_{\rm i}$ (metallicity) spans 0.007 -- 0.027 in steps of $\sim0.003$, while 
 $X_{\rm i}$ (hydrogen) is set to 0.70: this corresponds to an initial He
abundance  $Y_{\rm i} = 0.263 - 0.283$. 
 The mixing length parameter $\alpha = 2.0$ is used, which
 was obtained by calibrating it with solar data. 

To obtain the grid-based model stellar properties 
(\logg, \mass, \rad, \lum, \age)
we perturb the set of input observations using a random Gaussian
distribution, and compare the perturbed observations to the model ones.
The input observations consist primarily of \mlsep, \numax, \teff, and \feh, 
although other inputs are possible, for example, \lum\ or \rad.
The stellar parameters and  uncertainties are defined as the mean value of the 
fitted parameter from 10,000 realizations, with the standard deviations 
defining the 1$\sigma$ uncertainties.

\subsection{Analysis approach\label{sec:analysisapproach}}


We determine a seismic \logg\ for the stars using the method
explained above, and the following data sets:\\
(S1) \{\mlsep, \numax, \teff, \feh\}, \\
(S2) \{\mlsep, \numax, \teff \}, and \\
(S3) \{\mlsep, \numax \}.\\
For the potential sample of Gaia calibration stars, \feh\ is not always
available, and in some cases, a photometric \teff\ may have various estimations.
For these reasons we include S2 and S3.

The observational errors in our sample are very small due to the 
brightness and proximity of the star, so we also derive an 
asteroseismic \logg\ while considering observational errors that we expect
for \kep\ stars (see \citealt{ver11}). 
We consider three types of observational errors:\\
(E1) the true measurement errors from the literature,\\
(E2) typically ``good'' errors, 
i.e. $\sigma($\mlsep,\numax,\teff,\feh) = 0.5 \mhz, 5\%, 70 K, 0.08 dex,\\
(E3) ``not-so-good'' errors (e.g.
V$\sim$11,12), 
$\sigma($\mlsep,\numax,\teff,\feh) = 1.1 \mhz, 8\%, 110 K, 0.12 dex.

\subsection{Seismic versus direct \logg\label{sec:seismicdirect}}
In Figure \ref{fig:logg} we compare the asteroseismic \logg\ with the 
true \logg\ for the seven stars.
Each star is represented by a point on the abscissa, and the y-axis
shows (seismic - true) value of \logg.
There are {three} panels which represent the results using the three
different
subsets of input data. 
We also show for each star in each panel three results; in the bottom left 
corner these are marked by 'E1', 'E2', and 'E3', and  
represent the results using the different errors in the observations.
The black dotted lines represent (seismic - true) \logg\ = 0, and the grey
dotted lines indicate \pmm 0.01 dex.

\begin{figure}
\centering
\includegraphics[width=0.56\textwidth]{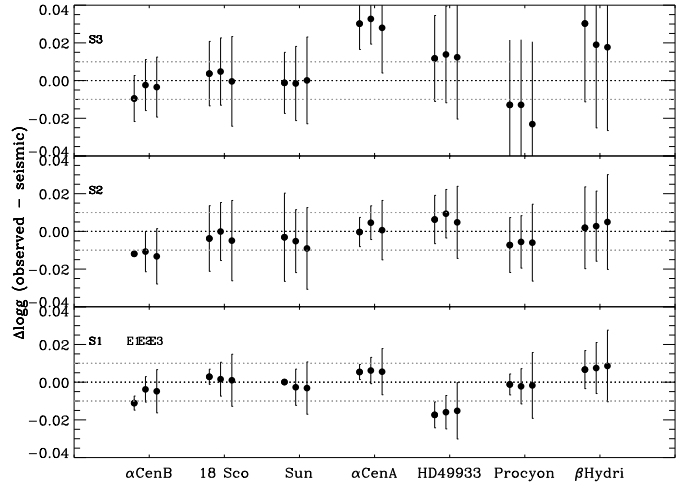}
\caption{{\it Seismic-minus-true} \logg\ for the seven sample stars 
while considering different sets of input observations (different panels)
and different observational errors (E1, E2, E3).\label{fig:logg}}
\end{figure}

Figure~\ref{fig:logg} shows that for all observational sets and errors
\logg\ is generally estimated to within 0.02 dex in both precision and 
accuracy.
This result clearly shows the validity of the mean seismic quantities and 
atmospheric parameters for providing an extremely precise value of \logg.
Other general trends that can be seen are 
that the typical {\it precision}
in \logg\ decreases as (1) the observational errors increase (from E1 -- E3),
and (2) the information content decreases (S1 -- S2 -- S3, for example).
One noticeable result is the systematic offset in the derivation of \logg\ for
HD~49933 when we use \feh\ as input (S1).
This could be due to an incorrect metallicity, an error in the adopted 
true \logg\ or a shortcoming of the grid of models.

\begin{table*}
\begin{center}
\caption{Stellar properties for the reference stars derived
by RadEx10\label{tab:radexgmr}}
\begin{tabular}{lccccccccccccccccccc}
\hline\hline
Star &  \logg (dex) & $R$ (\rsol) & $M$ (\msol) & $L$ (L$_{\odot}$) & Age (Gyr)\\
\hline
$\alpha$ Cen B & 4.527 \pmm\ 0.004 & 0.859 \pmm\ 0.007 & 0.905 \pmm\ 0.023 &  0.52 \pmm\  0.02 &  9.4 \pmm\  2.0\\
18 Sco & 4.441 \pmm\ 0.004 & 1.018 \pmm\ 0.008 & 1.042 \pmm\ 0.019 &  1.07 \pmm\  0.04 &  4.8 \pmm\  0.9\\
Sun & 4.438 \pmm\ 0.001 & 1.000 \pmm\ 0.002 & 1.000 \pmm\ 0.005 &  1.01 \pmm\  0.03 &  6.3 \pmm\  0.6\\
$\alpha$ Cen A & 4.312 \pmm\ 0.004 & 1.223 \pmm\ 0.010 & 1.119 \pmm\ 0.024 &  1.56 \pmm\  0.08 &  7.0 \pmm\  0.9\\
HD\,49933 & 4.195 \pmm\ 0.007 & 1.418 \pmm\ 0.022 & 1.148 \pmm\ 0.054 &  3.23 \pmm\  0.22 &  3.5 \pmm\  0.6\\
Procyon & 3.981 \pmm\ 0.006 & 2.072 \pmm\ 0.024 & 1.497 \pmm\ 0.041 &  7.08 \pmm\  0.54 &  2.1 \pmm\  0.2\\
$\beta$ Hydri & 3.957 \pmm\ 0.010 & 1.840 \pmm\ 0.045 & 1.119 \pmm\ 0.086 &  3.55 \pmm\  0.31 &  6.8 \pmm\  1.0\\

\hline
\end{tabular}
\end{center}
\end{table*}

Figure~\ref{fig:radmasss0} shows the {\it seismic} 
radius and mass determinations
of the sample stars using S1 while considering the three sets of errors.
We find that with  {\it good} observational errors, the radii are matched
to within 1\% (accuracy) with typical precisions of 2--3\%, while the masses 
are matched to within 1--4\% with typical precisions of 4--7\%.
Here it can be seen that the offset found for HD~49933 in \logg\ is 
related to the reference mass value (the radius seems to be consistent).
For S2 and S3 the uncertainties begin
to grow very large; 2--5\% and 3--10\%, respectively in radius, 
and 5--15\% and 10-35\% in mass, while 
the accuracies also decrease (not shown) although to within $<1.5\sigma$ for 
all results.
These results indicate that for the most precise determination of 
mass and radius, a seismic index and both \teff\ and \feh\ are necessary,
unlike \logg\ where the seismic information alone (or including \teff)
can produce an accurate result.


\begin{figure}
\includegraphics[width=0.45\textwidth]{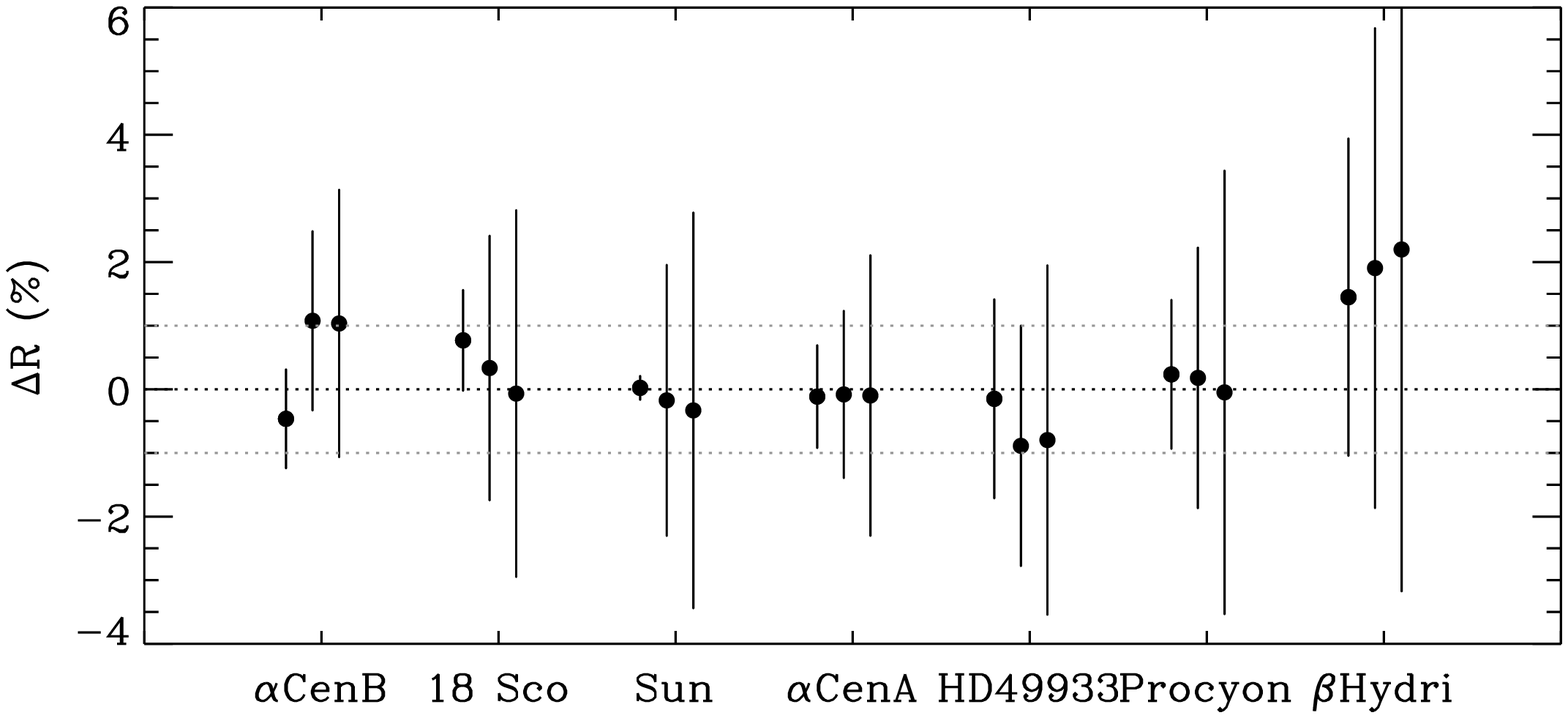}
\includegraphics[width=0.45\textwidth]{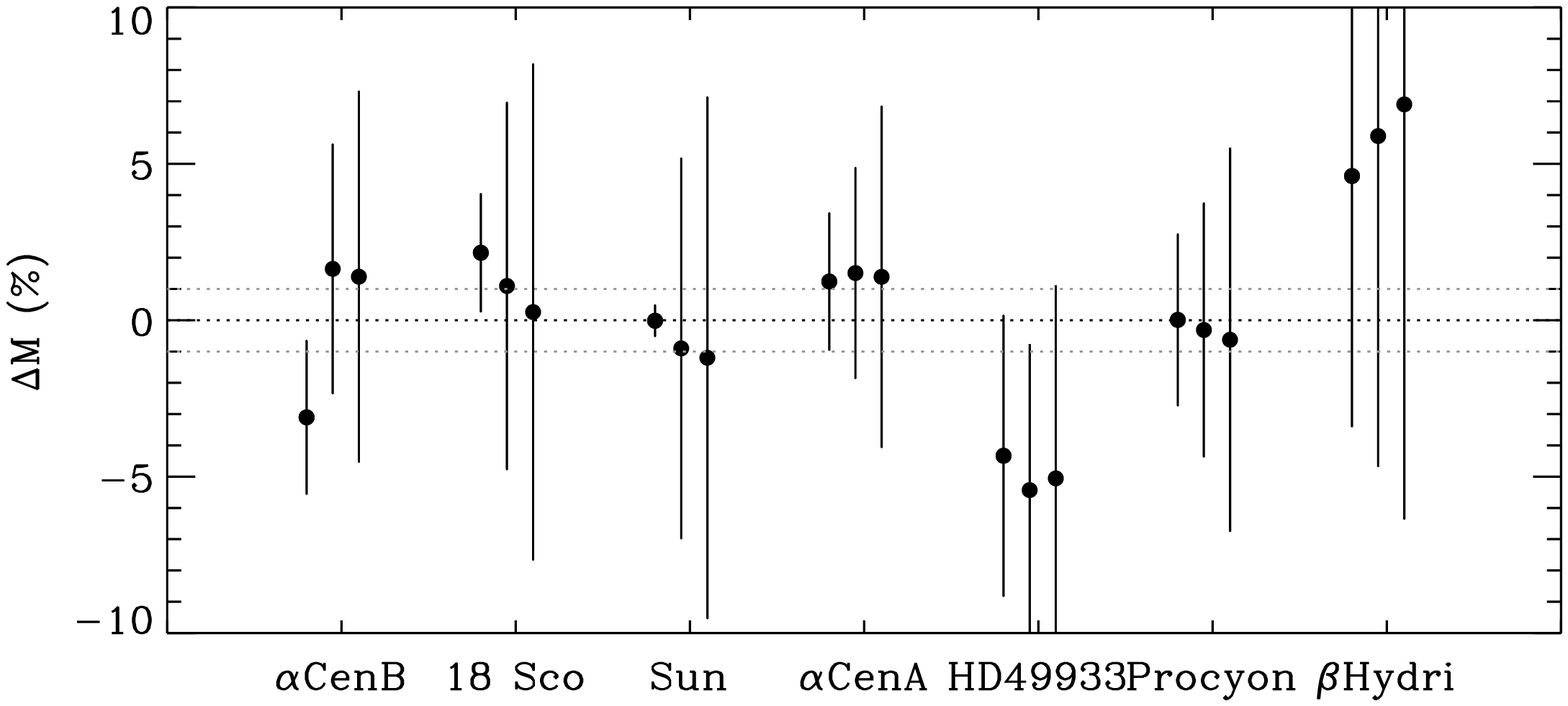}
\caption{{\it Seismic-minus-True} values of radius (left panel) and 
mass (right panel) using S1.{\label{fig:radmasss0}}}
\end{figure}

\subsection{Systematic errors in observations}


To study the effect of {\it systematic errors} in the atmospheric 
parameters, we 
repeated our analysis for $\beta$ Hydri using three sets of input
data that change only in \teff\ and \feh\ considering the E2 errors.  
The first set (1) uses the \citet{nor07} values
(5872, --0.10), the second set (2) uses (5964, --0.10), 
and the third set (3) uses \citet{das06} values (5964, --0.03).
For S1 we derived \logg\ = 3.96, 3.97, and 3.97  dex
for case 1, 2, and 3, respectively (ref. value is 3.95 dex).
Excluding \feh\ (S2) we derived \logg\ = 3.95 and 3.96
for case 1 and 2, respectively.
Here we can conclude that errors in the atmospheric parameters can
change \logg\ by up to 0.02 dex, and in the absence of an accurate \feh\ it
is better to exclude it.   
To determine the mass, radius, and age, however, \feh\ is a very imporant 
constraint.

\section{Conclusions}

We summarize the stellar properties of the sample stars in 
Table~\ref{tab:radexgmr}
derived
by RadEx10 using \mlsep, \numax, \teff, and \feh, and the true observational
errors.  
We highlight the excellent agreement between seismically determined
parameters and those obtained by direct mass and radius estimates
(compare Tables~\ref{tab:refstars} and \ref{tab:radexgmr}). 
In only 
two cases ($\alpha$ Cen B and 18 Sco), we find that \logg\ and mass are determined 
with a difference of just over 1$\sigma$ for S1 and S3,
while for S2 we find that \logg\ is accurate to within its $\sigma$ for all stars.
This study validates the accuracy of a seismically determined \logg\ while
also highlighting the excellent precision that can be obtained using seismic data.
If we relax the observational errors to those typical of what is available
for the sample of $\sim$500 \kep\ F, G, K IV/V \kep\ stars, then we obtain
\logg\ with precisions of less than 0.02 dex for S1 (including \feh\ as a measurement)
and less than 0.03 dex for S2 (excluding \feh) for even ``poor'' observational errors
on the input seismic and atmospheric data.
We also showed that we 
can expect to find a typical systematic error of no bigger than 0.02 dex
arising from an error in the atmospheric parameters.


\begin{acknowledgements}
OLC is a Henri Poincar\'e Fellow at OCA, funded by the Conseil 
G\'en\'eral des Alpes-Maritimes and OCA.
\end{acknowledgements}

\bibliographystyle{aa}  
\bibliography{ref} 

\end{document}